\documentclass[preprint]{aastex}
\shorttitle{PIAA for the centrally obscured Subaru Telescope pupil}
\shortauthors{J. Lozi, F. Martinache, O. Guyon}
\usepackage{graphicx}
\usepackage[mathscr]{eucal}
\begin{document}

\title{
  Phase-Induced Amplitude Apodization on centrally obscured pupils:
  design and first laboratory demonstration for the Subaru Telescope
  pupil
}

\author{
  Julien Lozi\altaffilmark{1},
  Frantz Martinache\altaffilmark{1} and  
  Olivier Guyon\altaffilmark{1}.
}

\altaffiltext{1}{National Astronomical Observatory of Japan, 
  Subaru Telescope, Hilo, HI 96720}
\email{frantz@naoj.org}

\begin{abstract}
High contrast coronagraphic imaging is challenging for telescopes
with central obstructions and thick spider vanes, such as the Subaru
Telescope.
We present in this paper the first laboratory demonstration of a high
efficiency PIAA-type coronagraph on such a pupil, using coronagraphic
optics which will be part of the Subaru Coronagraphic Extreme-AO
(SCExAO) system currently under assembly. Lossless pupil apodization
is performed by a set of aspheric PIAA lenses specifically designed to
also remove the pupil's central obstruction, coupled with a Spider
Removal Plate (SRP) which removes spider vanes by translating four
parts of the pupil with tilted plane-parallel plates. An
"inverse-PIAA" system, located after the coronagraphic focal plane
mask, is used to remove off-axis aberrations and deliver a wide field
of view.

Our results validate the concept adopted for the SCExAO system, and
show that the Subaru Telescope pupil can properly be apodized for high
contrast coronagraphic imaging as close as $\approx$ 1 $\lambda/D$
with no loss of sensitivity. We also verify that off-axis aberrations
in the system are in agreement with theory, and that the inverse PIAA
system recovers a wide usable field of view for exoplanet detection
and disks imaging.
\end{abstract}

\section{Introduction}

While the existence of large numbers of extrasolar planets around
solar type stars has been unambiguously demonstrated by radial
velocity (RV), transit and microlensing surveys, attempts at their direct
imaging with AO-equipped large telescopes remain largely unsuccessful.
Due to modest AO performance and lack of high performance coronagraphs, 
the planet-rich inner parts of solar systems (including the habitable
zones) are still out of the reach of current imaging surveys, which
are only sensitive to planets beyond $\sim0.2\arcsec$. Several recent
surveys \citep{2007ApJ...670.1367L, 2007ApJS..173..143B,
  2007A&A...472..321K} indicate that massive planets on such wide
orbits are rare, although recent observations have found candidates
around the young A-stars HR 8799 \citep{2008arXiv0811.2606M}, $\beta$
Pictoris \citep{2008arXiv0811.3583L} and Fomalhaut
\citep{2008Sci...322.1345K}.

Many coronagraph concepts have been developped over the last ten
years, most of them derived from Bernard Lyot's original design
\citep{1939MNRAS..99..580L}. While most designs theoretically offer
very high contrast, the performance is in practice severely
compromised by residual wavefront errors after correction by the AO
system \citep{1999PASP..111..587R, 2005ApJ...629..592G,
  2006A&A...447..397C} and by the optical layout of telescopes,
especially the presence of a central obscuration and spider vanes due
to the secondary mirror \citep{2005ApJ...633..528S,
  2008A&A...492..289M}.

Yet a new generation of high-contrast imaging instruments is coming
online: GPI on Gemini South \citep{2008SPIE.7015E..31M}, SPHERE on VLT
\citep{2006Msngr.125...29B} and SCExAO/HiCIAO on Subaru
\citep{2008SPIE.7014E..42H}. They all use so-called extreme AO
systems which employ deformable mirrors (DM) with large number of fast
actuators and new wavefront control techniques to provide the best
raw contrast ($\sim 10^{-5}$) accessible from the ground in the near
infrared.

In the first phase of the SCExAO project (spring 2010), a high
performance PIAA coronagraph will be implemented with a 1024-actuator
MEMs-based deformable mirror (DM) as an upgrade that will feed
Subaru's newly commissioned coronagraphic imager instrument HiCIAO
\citep{2008SPIE.7014E..42H} in the context of the Subaru Strategic
Exploration of Exoplanets and Disks (SEEDS) campain.
One of the key scientific drivers of HiCIAO is to use direct imaging
to investigate the presence of Jovian-mass planets around young
stars. Current HiCIAO observations in Angular Differential Imaging
(ADI) mode using Subaru's AO system (AO188) can only efficiently probe
for companions at angular separations greater than $\sim$0.5\arcsec and
reach maximum sensitivity around 1 \arcsec. For instance,
\citet{2007ApJ...670.1367L} respectively report H-band contrasts
sensitivities of 9.5 and 12.9 magnitudes for these separations.
SCExAO was designed to complement this search parameter space and
allow to probe the innermost parts of extrasolar planetary systems. It
is essentially an optical bench that will replace HiCIAO's current
fore-optics (standard Lyot coronagraph), for follow-up observations of
challenging targets that would benefit from better AO corrections.

At a given angular separation, the achieveable raw contrast level in
an Extreme-AO system is mostly driven by the speed and accuracy of the
wavefront control system. The coronagraphic solution we report in this
paper and adopted for SCExAO was designed to achieve a $10^{-6}$ raw
contrast at 1 $\lambda/D$ angular separation in the absence of
wavefront aberrations at the input of SCExAO. The system is optimized
for the H-band (1.6 $\mu$m).

It the first installment of the SCExAO bench, the SCExAO DM will be
used to calibrate the slowly varying aberrations in order to reduce
the level of quasi-static and slow moving speckles \citep{guyon09}.
Fast correction (1 kHz correction frequency) will be provided by the
188-actuator Subaru curvature system ahead of SCExAO.
A subsequent upgrade of the system will include an internal fast
visible wavefront sensor that will feed the SCExAO DM and actually run
in extreme AO mode.
The raw contrast is then expected to improve to $10^{-4}$ (detailed
performance estimates are indicative rather than predictive, and
subject to performance validation in a laboratory prototype currently
under design). The $100\times$ to $1000\times$ gain between raw
contrast and detection limit will be achieved through time averaging
of speckles in the focal plane, augmented by calibration and
differential imaging techniques. Consequently, the coronagraph system
must be designed so that it does not create fixed speckles much above
the $10^{-6}$ contrast level.

While static phase errors in the coronagraph can be removed by the DM,
speckles or diffraction features due to amplitude errors are harder to
remove by using phase on the DM: they can require large stroke and can
only removed from one side of the focal plane in a relatively narrow
spectral bandwidth. The SCExAO system is therefore designed to
minimize such features by amplitude apodization of the pupil
(including removing the large central obstruction) with a Phase
Induced Amplitude Apodization (PIAA) \citep{2003A&A...404..379G,
2005ApJ...622..744G} and removal of the pupil spider vanes by a custom
Spider Removal Plate (SRP).
This paper shows that these techniques do not compromize throughput or
angular resolution, and allow high contrast imaging as close as 1
$\lambda/D$ from the optical axis.

We introduce in \S \ref{sec:apodize} the design of these two
key optical components of the SCExAO system and present laboratory
results (\S \ref{sec:lab}) that validate their basic functionality.
We also show that that the static aberrations these components
introduce are within the correction range of the SCExAO deformable
mirror, and will therefore not affect the overall SCExAO performance.
Finally, our laboratory demonstration shows how the pupil remapping
distorts images of off-axis sources (\S \ref{sec:offaxis}). We
however demonstrate that these aberrations are well compensated by an
inverse PIAA.

The tests reported in this paper were conducted with no wavefront
control system, and we therefore provide no measurement of contrast
performance for the SCExAO system at this time. The implementation of
a wavefront control system within SCExAO will be the next step of
development for this project. High contrast results with an integrated
PIAA coronagraph + wavefront control system are reported in a separate
paper \citep{guyon09}.

\section{Lossless apodization of a centrally obscured beam}
\label{sec:apodize}

With a central obscuration and spider vanes, the pupil of large ground
based telescope is incompatible with most high performance coronagraph
concepts, although a few notable exceptions exist (see for example
\citet{1997PASP..109..815R,2005ApJ...618L.161S}).

On the Subaru Telescope pupil, the size of the central obstruction (30
\%, linear) and spider thickness (22cm) require that the coronagraph
is designed to remove the diffraction features they create in the
focal plane. The spiders, if left uncorrected, create 4 spikes at
$\sim 10^{-3}$ contrast. These spikes are especially problematic at small
angular separation, where their cover most of the position angle
space. The central obscuration, at 30 \%, creates its own set of
diffraction rings with a $10^{-3}$ contrast level at the peak of the
central obstruction's first ring (approximately 10 $\lambda/D$). Any
coronagraph which is designed to offer contrast better than $10^{-3}$
on Subaru Telescope therefore needs to take into account the spiders
and central obstruction. We note that several coronagraph concepts can
be designed to mitigate the effects of such features. For example, in
Lyot type coronagraphs, the pupil Lyot plane can mask most of the
diffraction due to the central obstruction and spiders. This is
usually achieved at the expense of coronagraph throughput (for
example, a larger fraction of the pupil is masked by the Lyot mask).
\citet{2008A&A...492..289M} show that this option is suitable for a
contrast goal of $10^{-5}$, a spider thickness equal to 1.8 \% of the
pupil diameter and a coronagraph with a 2.4 $\lambda/D$ inner working
angle.
We have not carefully evaluated this option for SCExAO, but we note
that it may be difficult to implement given the large spider thickness
(2.8 \% of the telescope diameter) on Subaru, and the SCExAO goal to
keep diffraction features due to spiders at or below $10^{-6}$
contrast level at 1 $\lambda/D$ from the optical axis.

The small inner working angle of the SCExAO coronagraph makes this
Lyot mask approach much less efficient, as the small focal plane mask
diameter creates a correspondingly large halo around the spiders ($2.4
\times$ larger than for a 2.4 $\lambda/D$ inner working angle
coronagraph), so a larger area would need to be masked by the Lyot
stop.
A variant, proposed by \citet{2006A&A...451..363A} consists in
inserting a complex amplitude filter in the pupil Lyot plane designed
to produce a fully cleared output pupil, which can improve the throughput.

In this paper, we demonstrate the feasibility of a different approach:
our coronagraph is designed to remove (or at least greatly reduce)
both the spiders and central obstruction by geometrically remapping
the pupil.
Indeed, the classical apodization technique used on most coronagraphs
reduces the effective diameter of the telescope, therefore affecting
the inner working angle (to $\sim 4 \lambda/D$) as well as the overall
throughput. SCExAO uses a PIAA-based coronagraph, which preserves both
qualities. Compared to classical apodization, this ``boost'' in
resolution allows to probe for the presence of companions at angular
separations close to one $\lambda/D$.

First, a device based on geometrical optics removes the spider vanes
by translating the four "pupil quadrants" they define toward the
center of the pupil. This device, called the Spider Removal Plate
(SRP), as well as its manufacturing are introduced in details in
\S\ref{sec:SRP}. 
The beam is then apodized using a Hybrid Phase Induced Amplitude
Apodization scheme \citep{2006ApJ...644.1246P}, compatible with the
presence of a central obscuration. The apodizer is presented in
\S\ref{sec:PIAA}.

\subsection{A plate to remove the spider vanes}
\label{sec:SRP}

\begin{figure}
  \plotone{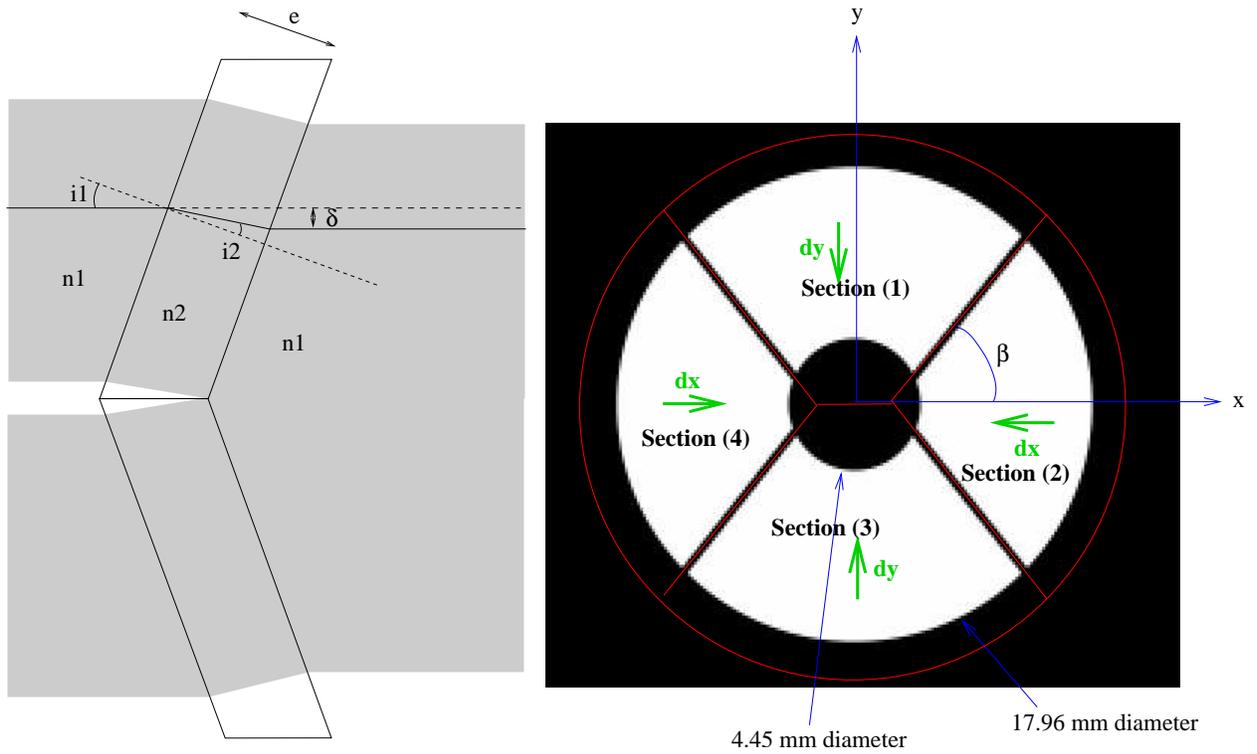}
  \caption{
    Left panel: cross-section of the Spider Removal Plate. The
    translation of the beam $\delta$ is a function of the tilt angle
    $i_1$ as well as the index and the thickness $e$ of the window.
    Right panel: geometry of the Subaru Telescope pupil. The SRP
    translates quadrants 1 and 3 inward by the amount $\delta$y
    and quadrants 2 and 4 by $\delta$x.}
  \label{fig:SRP}
\end{figure}

Our approach, illustrated in Figure \ref{fig:SRP}, is to translate
each of the four parts of the beam with a single tilted plate of
glass, in a manner reminiscent of the technique of pupil densification
\citep{1996A&AS..118..517L}. The translations are computed to fill the
gap due to the spiders.
The spider vanes of the Subaru Telescope are 224 mm thick, for a total
pupil diameter of 7.92 m. To account for alignment errors between the
SRP and the pupil, we have chosen to design the SRP for slightly
thicker (250 mm) spider vanes.
The coronagraph we are assembling for Subaru is designed for an input
beam of 17.96 mm in diameter. At this scale, the gap due to the spider
vanes is therefore 0.453 mm.
The SRP consists of four tilted plane-parallel plates, each
translating a part of the pupil inwards, as shown in Figure
\ref{fig:SRP}. It can be best described as a ``pyramid-shaped
rooftop'' of constant thickness. 
All four plates were cut from the same same plane-parallel plate
(a.k.a. optical window), to guarantee, within tolerances, a constant
thickness.

Commercially available optical windows have, for typical
specifications, a parallelism better than five arc seconds, that is
about 2.5 $10^{-5}$ radians. Across a 1 cm long piece cut from such a
window, the maximum wedge is going to be 0.25 microns ($\sim\lambda/7$
in H-band) which is well within the comfort zone of a deformable
mirror. All four plates of or custom component were cut from a single
window of thickness $e = 15$ mm and parallelism better than three
arc seconds and the geometry of the cuts and assembly was optimized to
avoid pasting together areas which originated from distant parts of
the original window.

To translate a quadrant, each of the four segments of the SRP is
tilted by an angle $\alpha$ that is a function of the amplitude of the
desired translation. One respectively calls $\delta x$ and $\delta y$
the horizontal and vertical translation of the pupil quadrants (cf. 
Fig. \ref{fig:SRP}), and $\beta$ the angle between the horizontal axis $x$
and the top right vane of the Subaru Telescope pupil. 
The goal of the SRP beeing to fully fill the spider gap $\gamma$ =
0.453 mm, $\delta x$ and $\delta y$ must verify:

\begin{equation}
\delta x \sin\beta + \delta y \cos\beta = \gamma.
\label{eq:srp}
\end{equation}

Tilting a SRP segment by an angle $\alpha$ not only displaces the
corresponding quadrant by $\delta$, it also introduces an optical path
difference (OPD) $\Delta$ relative to a hypothetical undeflected beam.
The relation between the tilt angle $\alpha$ (identical to $i_1$ in
Fig. \ref{fig:SRP}), the deplacement $\delta$ and the OPD $\Delta$
come from equations of geometric optics (cf. left panel of
Fig. \ref{fig:SRP}):

\begin{eqnarray*}
\sin i_{1} &=& n \sin i_{2} \textrm{ and} \label{eq:snell}\\
\delta &=& e \sin(i_{1}-i_{2})/\cos i_{2} \label{eq:delta}, \\
\Delta &=& \frac{e}{\cos i_{2}}(n-\cos(i_{1}-i_{2})) \label{eq:opd}.
\end{eqnarray*}

\noindent
where $n$ is the refractive index of the glass (index of air
considered to be exactly one). In the small angle approximation, these
relations reduce to:

\begin{eqnarray}
\delta &\approx& e \alpha (n-1)/n \label{eq:smalldelta} \\
\Delta &\approx& e (n-1) (1+ \alpha^2/2n). \label{eq:opd:dl}
\end{eqnarray}

\begin{figure}
  \plotone{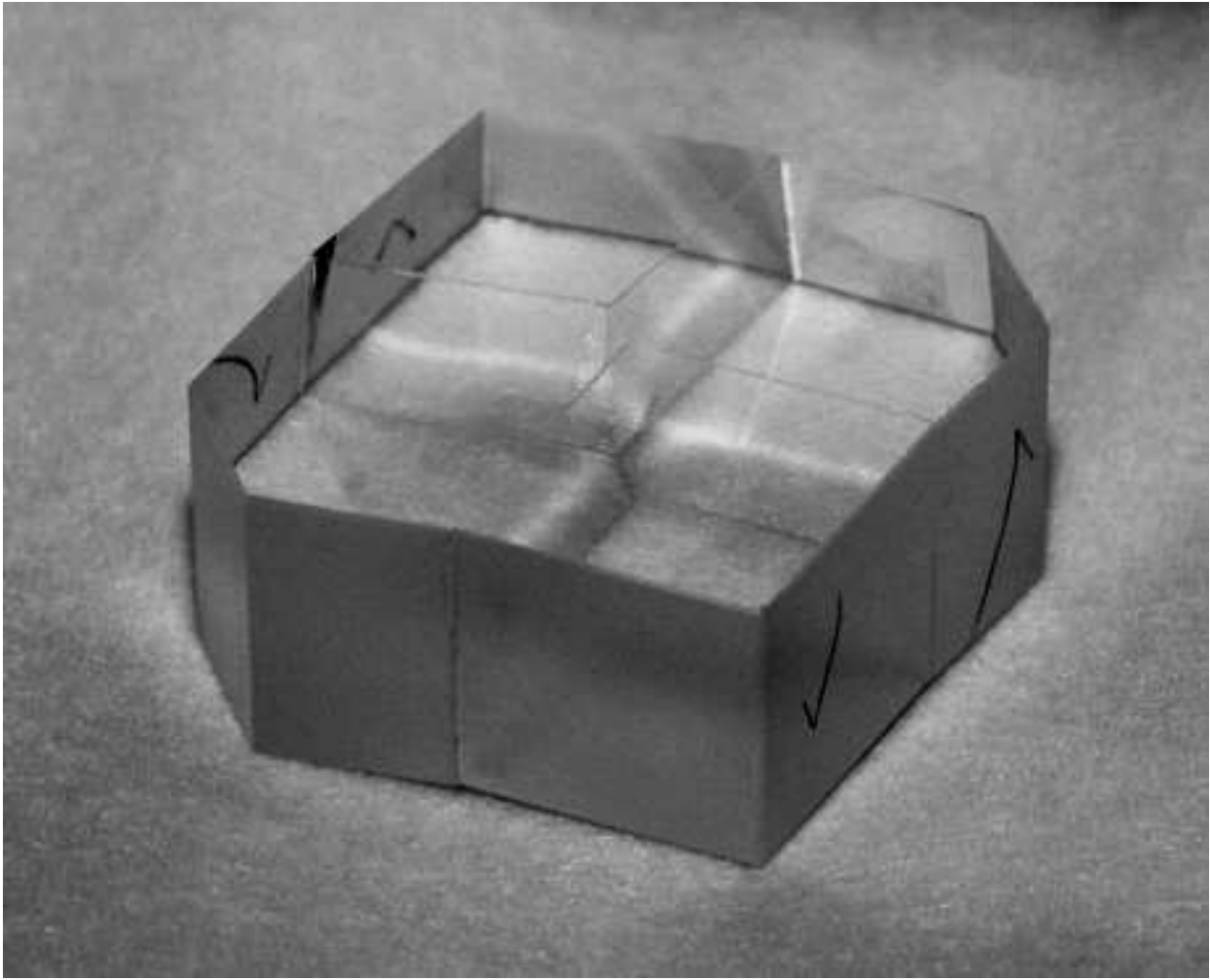}
  \caption{Picture of the assembled Spider Removal Plate. The plate is
    15 mm thick. Note that since the spider vane angle $\beta$ is
    $51.75^\circ$ and not $45^\circ$, the SRP is not continuous at the
    interfaces between the four plates and there are steps running
    along them.}
  \label{fig:photoSRP} 
\end{figure}

For the SRP to be useful, one must ensure that the OPD is the same
for all four quadrants at all wavelengths, so that the wavefront of an
on-axis source remains continuous at all wavelengths after the remapping.
This can only be achieved when the four plates have simultaneously the
same tilt angle $\alpha$. The solution to Eq. \ref{eq:srp} is
therefore to have $\delta x$ = $\delta y$ = $\delta$.

For a window of thickness $e=15$ mm and index $n=1.443$ (Fused Silica
for $\lambda$ = 1.6 $\mu$m), each plate needs to be tilted by an angle
$\alpha = 5.004^\circ$ (cf. Eq. \ref{eq:smalldelta}).
To guarantee the continuity of the wavefront on-axis after remapping
within $\lambda/10$, one can derive Eq. \ref{eq:opd:dl} relative to
$\alpha$ and find that the tolerance for the OPD translates into a
constraint on the tilt angle of each plate:

\begin{equation}
  \sigma_\alpha = \frac{n \sigma_{\Delta}}{(n-1) e \alpha},
\end{equation}

\noindent
which gives a tolerance of 0.02$^\circ$ for the tilt angle. Fig. 
\ref{fig:photoSRP} shows a picture of the SRP that was assembled for
the SCExAO system.

The refractive index of fused silica beeing a monotonous decreasing
function of the wavelength, a simple way to ensure that the SRP fully
fills the spider vanes over the entire H-band (1.485 - 1.785 $\mu$m)
is to optimize it for the ``red side'' of the band which is less
deflected than the ``blue side''. The two main non-ideal effects of
the SRP are:

\begin{itemize}
\item The amplitude of the lateral shifts is chromatic, and the pupil
  size at the SRP output is therefore wavelength dependent. This
  effect is however small (0.3 \% over the entire H-band) and can be
  accounted for in the coronagraph design.
\item The output wavefront for off-axis sources contains ``steps''
  which produce aberrations. These aberrations limit the useful field
  of view and will be quantified in \S \ref{sec:SRP:fov}. One
  solution to this problem is to include a ``reverse-SRP'' after the
  coronagraph but before the science camera to recover a wide field of
  view.
\end{itemize}

After the SRP, the beam has no spiders, but still has a central
obstruction.

\subsection{PIAA with a central obscuration}
\label{sec:PIAA}

\begin{figure}
  \plotone{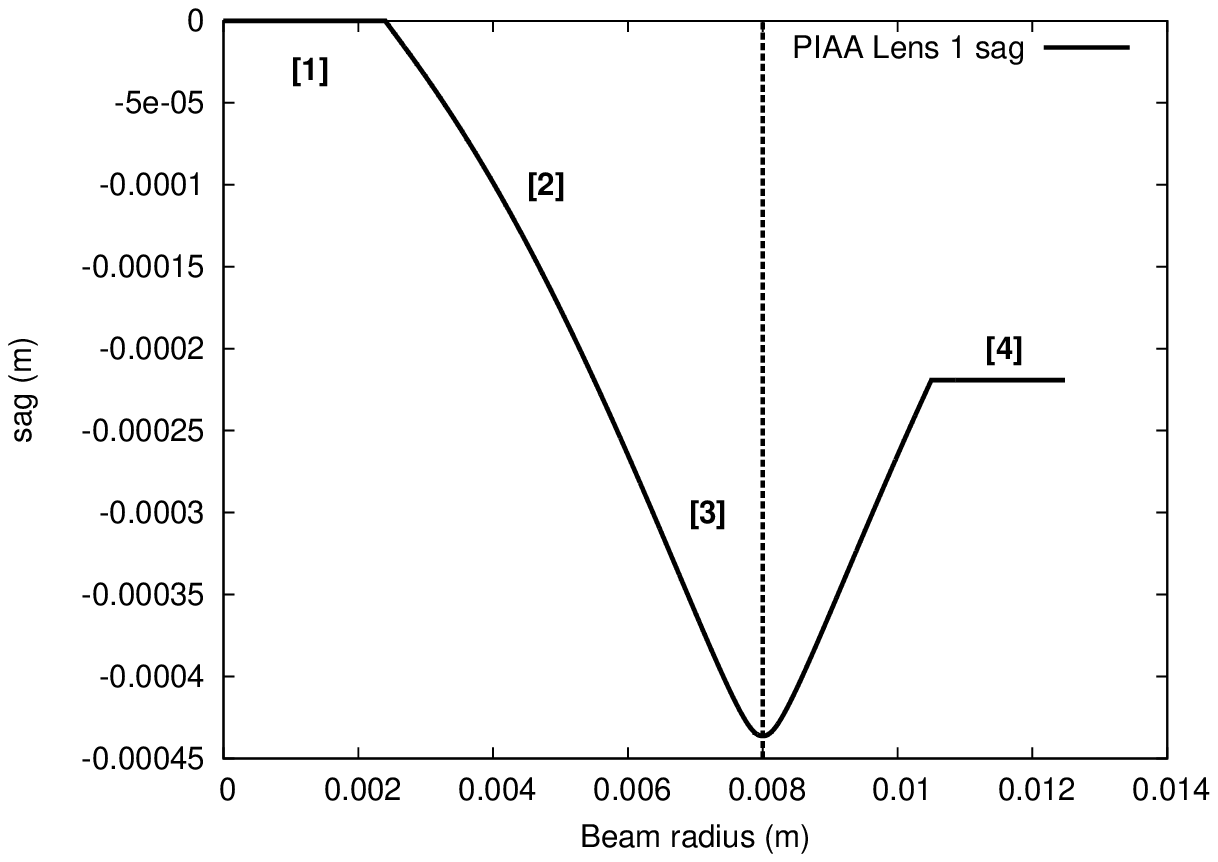}
  \plotone{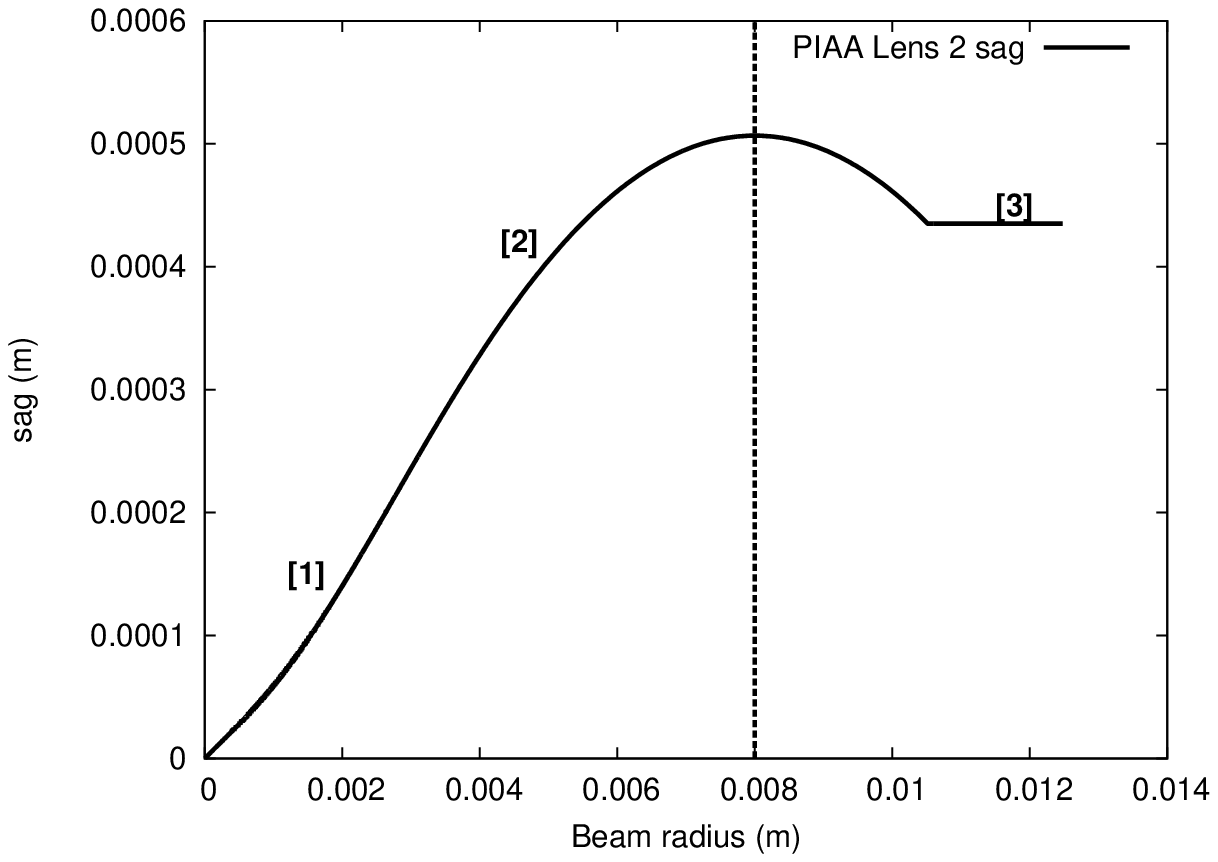}
  \caption{Radial sag profile for the two PIAA lenses. On both panels,
    the vertical dashed line localizes the outer edge of the pupil
    (radius 8 mm). Labels on the two curves refer to the description
    of the lenses' main features listed in the text of the paper.}
  \label{fig:lenssag} 
\end{figure}

Although not part of the original design of
\citet{1939MNRAS..99..580L}, apodization of the pupil has been
demonstrated to provide a major enhancement of most coronagraphs
\citep{2003EAS.....8...79A}.
The best apodizing function depends on the pupil geometry as well as
the type of coronagraph, but prolate spheroidal functions in general
have been shown to be natural apodizers for Lyot-type coronagraphs
\citep{2003A&A...397.1161S}, and solutions can be found for centrally
obscurated pupils \citep{2005ApJ...618L.161S}. 

The simplest way of performing an apodization is to insert a mask
whose radial transmission profile follows such a prolate spheroidal
function, the so-called classical pupil apodization (CPA). 
While extremely robust, and insensitive to moderate tip-tilt
residuals, this approach has two main drawbacks: the throughput is low
(as low as $\sim$ 0.1 for a $10^{-10}$ contrast
\citep{2003ApJ...582.1147K, 2006ApJS..167...81G}), and the effective
pupil diameter is reduced by a factor approximately equal to the
square root of the throughput (due to the fact that apodizers remove
the light mostly at the edges of the pupil), which translates into a
loss of angular resolution.

The Phase Induced Amplitude Apodization (PIAA)
\citep{2003A&A...404..379G}
addresses these issues and apodizes the beam using a very different
approach. It uses a set of two tailored optics working in pair:
inserted in the pupil plane, the first changes the distribution of
light, while the second collimates the beam for an on-axis source.
While this idea has already been studied in great details
\citep{2003A&A...404..379G, 2005ApJ...622..744G, 2006ApJ...639.1129M},
this paper presents a design adapted to the presence of a central
obscuration.

\begin{figure}
  \plotone{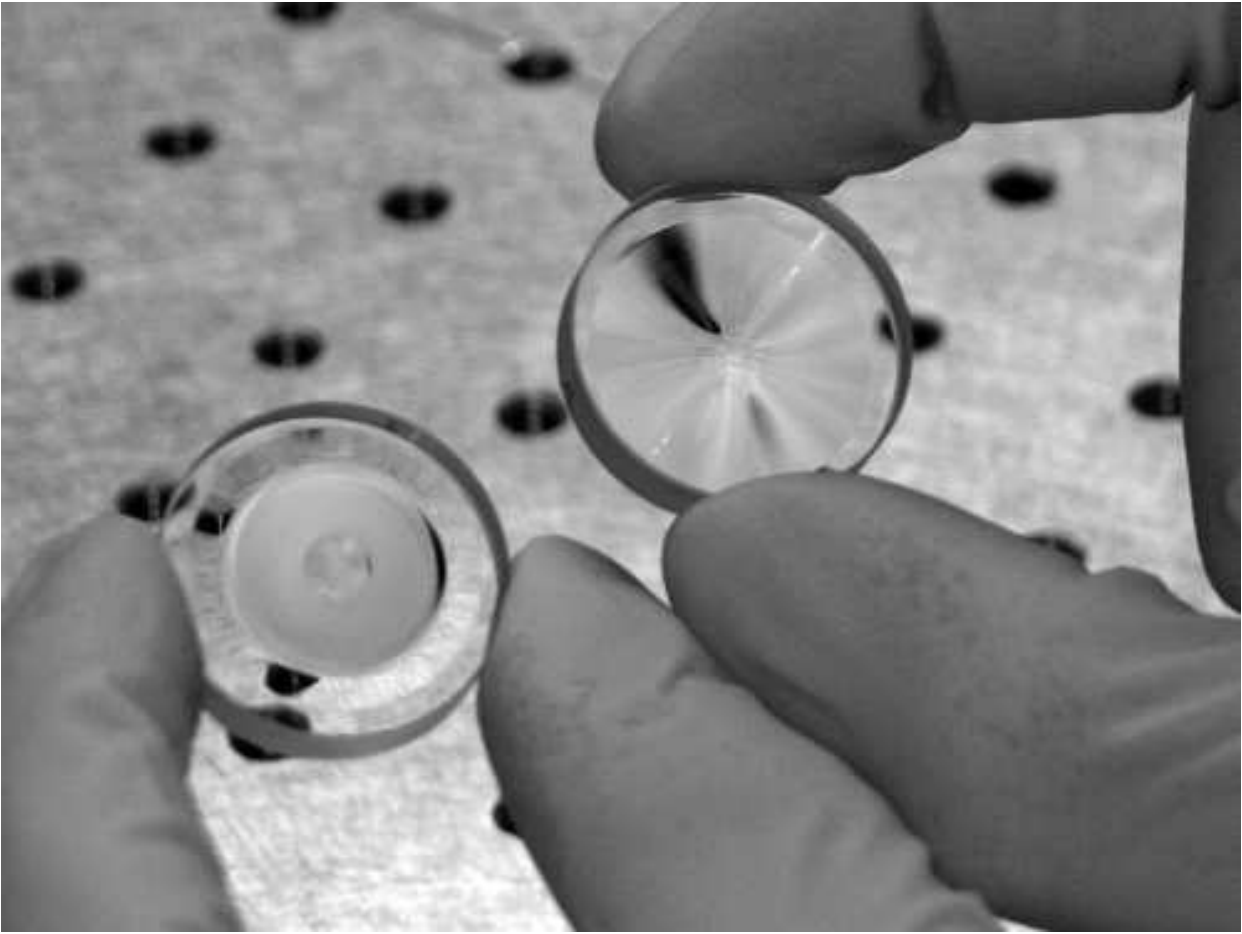}
  \caption{Picture of the PIAA lenses used to apodize the pupil of the
    Subaru telescope. L$_1$ (on the left) apodizes the beam and
    removes the central obstruction. L$_2$ (on the right) collimates
    the apodized beam.}
  \label{fig:PIAAlenses} 
\end{figure}

\citet{2006ApJ...644.1246P} have already shown that a PIAA/CPA hybrid
system is an attractive design that eases the manufacturability of the
first of the two PIAA optics near the outer edge of the pupil. This is
the design we have adopted for the SCExAO system.
These PIAA optics are designed as a refractive system (lenses instead
of mirrors). While such a choice would clearly be a risky option for a
space-based coronagraph aimed at reaching very high contrast
($\sim 10^{-9}$), refractive optics can safely be used in coronagraphs
for ground-based telescopes.

The set of PIAA lenses is an afocal system that leaves the pupil
diameter unchanged. A low dispersion material was chosen (CaF$_2$) to
keep chromatic aberrations small. Such a refractive PIAA system has
several advantages: its circular symetry makes it easier to
manufacture and it is quite compact ($96\pm 0.1$ mm long for a $\sim$
20 mm beam). Radial sag profiles of the PIAA lenses are presented in
Fig. \ref{fig:lenssag}. The profile for L$_1$ exhibits four main
features:

\begin{enumerate}
\item a central flat corresponding to what is left of the central
  obscuration after the SRP (cf. \S. \ref{sec:SRP}).
\item a convex (convergent) ring that densifies the inner part of the
  pupil and concentrates the distribution of light toward the center
  of the pupil.
\item a concave (divergent) ring that dilutes the outer part of the
  pupil. The telescope beam only illuminates L$_1$ out to 8 mm
  (vertical dashed line on Fig. \ref{fig:lenssag}), which corresponds
  to the bottom of the well, where the curvature radius is smallest.
\item an external flat ring that provides a convenient way of mounting
  the lens.
\end{enumerate}

The exact shape of the lens around the outer edge of the beam is
critical to the beam shaping, and greatly conditions the efficiency of
the downstream coronagraph. Yet because the curvature radius of the
lens is the smallest in that region, it is also where the biggest
manufacturing difficulty arises.

In order to relax the manufacturing tolerances, the PIAA was designed
to provide a partial apodization only, that will later be completed by
a binary mask (cf. \S \ref{sec:bin_mask}).
Also, while apparently superfluous, the right-hand side of the well in
the profile of L$_1$ acts as a safeguard making the system independent
of slight changes of scale in the pupil that typically accompany
realignment of upstream optics. Thus, an oversized pupil that would
otherwise induce light leaks sees its outermost part deflected away
from the beam, preserving the performance at an acceptable (a few
percents) cost in throughput.

The sag profile for L$_2$ is complementary of L$_1$ and exhibits three
main features:

\begin{enumerate}
\item a concave (divergent) central part that collimates the light
  that was densified by part $[2]$ of L$_1$.
\item a convex (convergent) ring that collimates the light that was
  remapped by part $[3]$ of L$_1$.
\item a flat external ring to facilitate mounting.
\end{enumerate}

Figure \ref{fig:PIAAlenses} shows a picture of the manufactured
PIAA lenses before beeing tested in the lab. The lenses are designed
to be separated by 96 mm exactly at $\lambda = 1.6 \mu$m, and adapted
for a 16mm-diameter beam.

\begin{figure}
\plotone{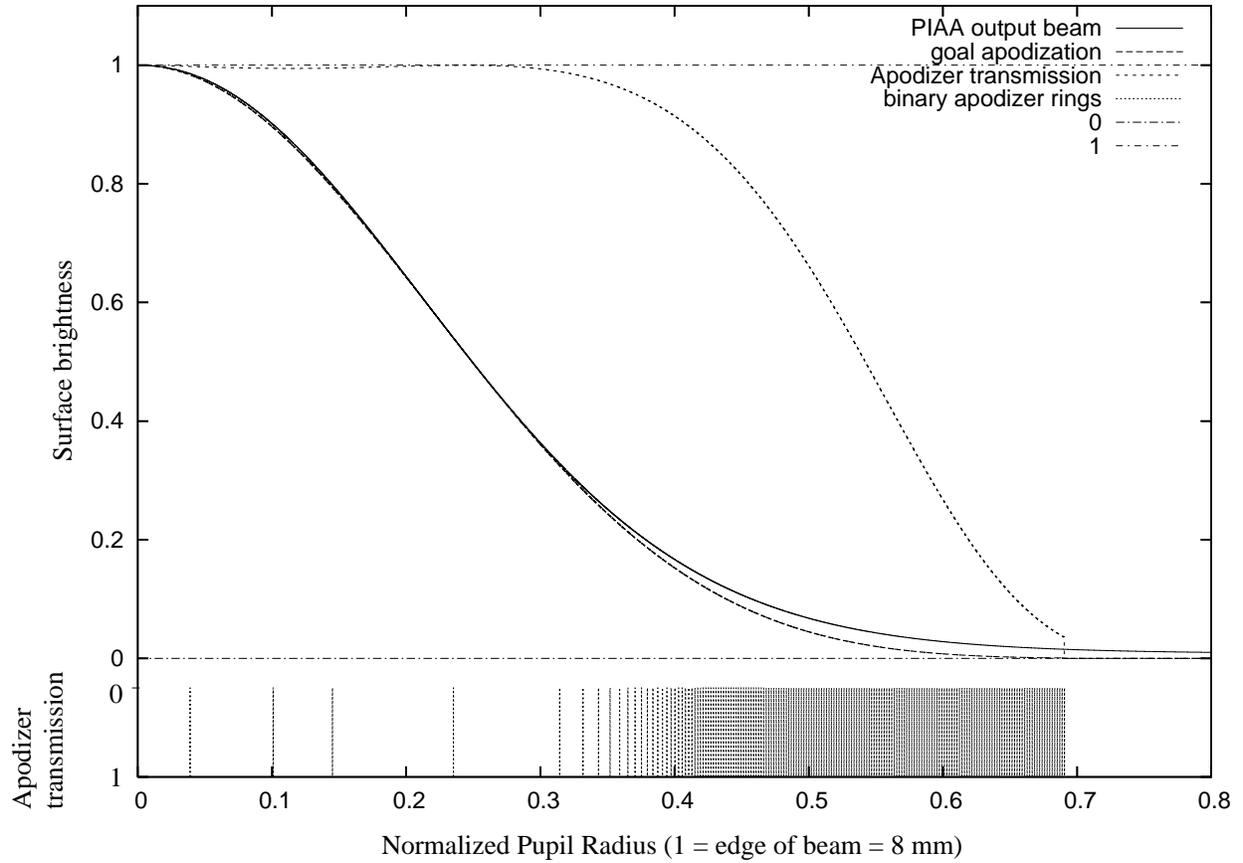}
  \caption{Binary mask transmission profile and discretization. The
  goal apodization (dashed line) is the result of the multiplication
  of the PIAA output beam profile (solid line) by the apodizer
  transmission profile (short dashed line).
  The rings in the binary apodizer (bottom) become very tightly spaced
  in the outer part of the beam, effectively blocking a larger
  fraction of the incoming light.}
  \label{fig:apodesign} 
\end{figure}

\subsection{Binary mask to complete the apodization}
\label{sec:bin_mask}

As explained in Section \ref{sec:PIAA}, the PIAA leaves a small
amount of light (0.82 \% of the central surface brightness) in the outer
part of the apodized pupil.
To be effective, the apodization therefore needs to be completed by a
CPA mask \citep{2006ApJ...644.1246P} in order to turn the (still) hard
outer edge of the pupil into a smooth transition. To perform this
complementary apodization, we chose to use a binary mask
\citep{2005ApOpt..44.1117K} with a series of narrow opaque rings
blocking light. With this approach, the apodizer is achromatic and its
transmission variations do not introduce phase variations on the
pupil.

\begin{figure}
  \plotone{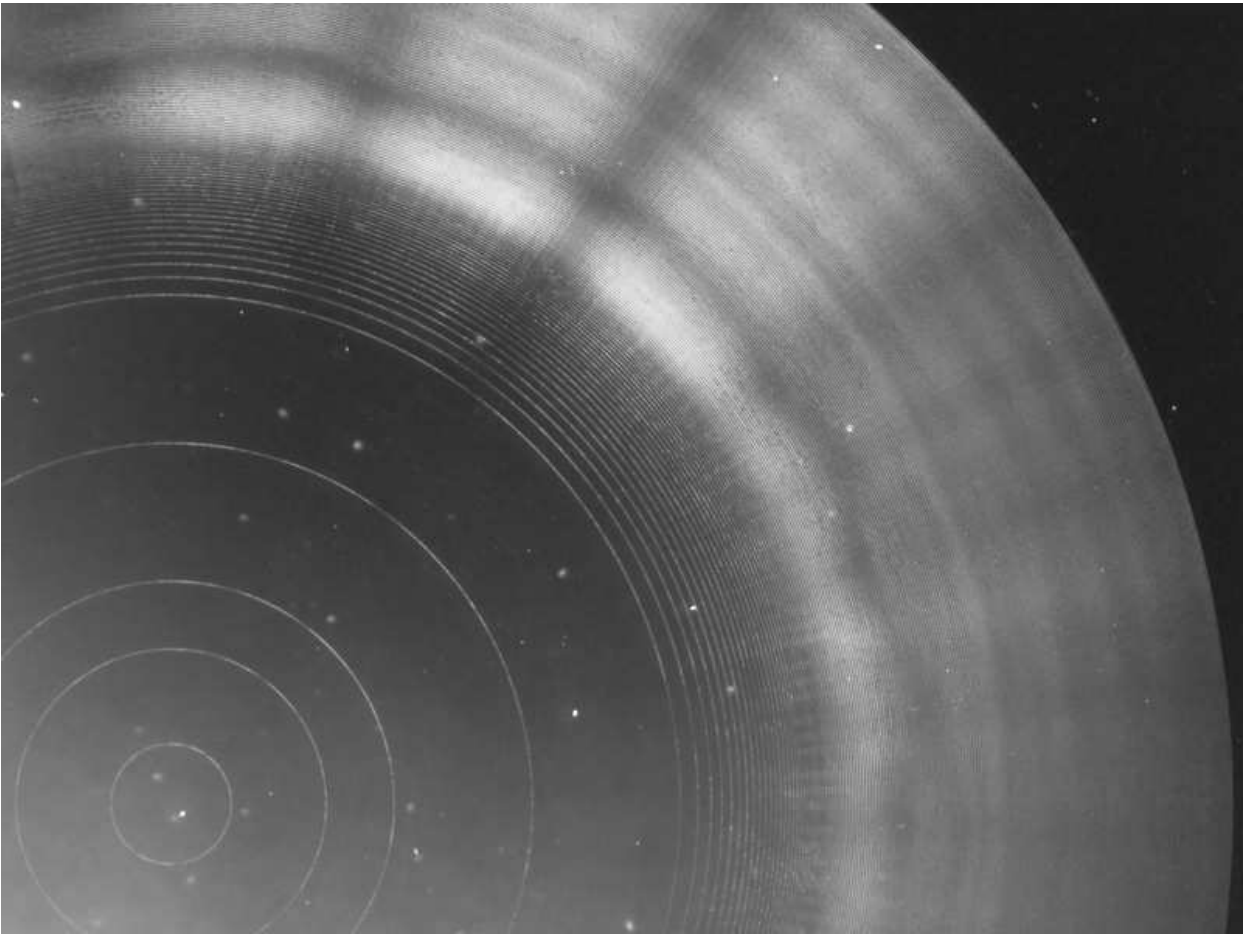}
  \caption{Picture of the binary mask taken with a
    microscope. The transmission, is about 10\% on the edge of the
    pupil.} 
  \label{fig:apo} 
\end{figure}

The position and width of the rings is optimized to best approximate
the ideal continuous apodization profile shown in Figure
\ref{fig:apodesign} as the ``Apodizer transmission'' curve. Several
constraints were imposed on the design to ensure manufacturability: no
ring should be less then 1.6 $\mu$m wide and the gap between
consecutive opaque rings should be no less than 15 $\mu$m. The
resulting design is composed of 155 opaque rings for a total apodizer
diameter of 11.09 mm (defined by the outer edge of the last opening
between opaque rings). The apodizer was manufactured by lithography on
a transmissive substrate. Figure \ref{fig:apo} shows a microscope
image of the mask.

The minimum width of the rings was limited to 1.6 $\mu$m to ease
manufacturing and avoid sub-wavelength features which could create
strong chromatic and polarization effects. We have not done a
diffraction analysis of the mask to quantify such effects, but we note
that (1) a similar design with 0.8 $\mu$m wide rings working in the
visible did not show significant deviation from Fraunhofer diffraction
theory \citep{guyon09} (2) the rings are mostly affecting the outer
part of the pupil where there is little incoming light.

The post-apodizer throughput over the 11.09mm diameter is 95.4\%, but
due to the narrow rings in the apodizer, some of the light transmitted
is diffracted at large angles. The ``effective'' throughput of the
apodizer is therefore lower, at 92.3\%, and would be equal to the
``raw'' throughput if the apodizer were continuous instead of
binary. An additional 4.8\% transmission is lost because the beam is
clipped by the apodizer at 69.3\% of the nominal 16 mm beam
diameter. This clipping is equivalent to clipping the outer 2.5\% of
the input telescope beam at the entrance of the PIAA system, and
relaxes pupil alignment tolerances at the expense of a 4.8\%
throughput loss and a 2.5\% angular resolution loss. The combined
throughput from (1) absorption in the apodizer, (2) diffractive losses
in the apodizer and (3) beam clipping is 87.5\%. We note that most of
these losses could be reduced with a more aggressive design with
tighter tolerances, but we decided for this first on-sky PIAA system
to keep the design more conservative.

\section{Laboratory testing}
\label{sec:lab}

\subsection{Optical layout}
\label{sec:layout}

Because of its specific design: incoming and outcoming beams are
collimated, and the aspheric surfaces face each other, the PIAA can
work over a broad range of wavelengths. The distance between the two
PIAA lenses can compensate for the change of refraction index when
changing wavelength. This allowed us to test the optics in visible
light, using a standard laser diode ($\lambda=633$ nm) as our light
source, although the system is optimized for the H-band ($\lambda =
1.6 \mu$m).

We have chosen to test the remapping optics (PIAA optics and SRP) with
visible light for convenience. The optics were designed for near-IR
wavelength, but their functionality can be tested with visible light
although near-IR optimized anti-reflection coatings create ghosts. We
note that a high contrast coronagraphic imaging test would require
near-IR light in addition to wavefront correction with a deformable
mirror, but in the absence of wavefront control, testing at a
wavelength almost three times shorter than nominal provides much
higher sensitivity to aberrations as well as alignment errors.

\begin{figure}
\plotone{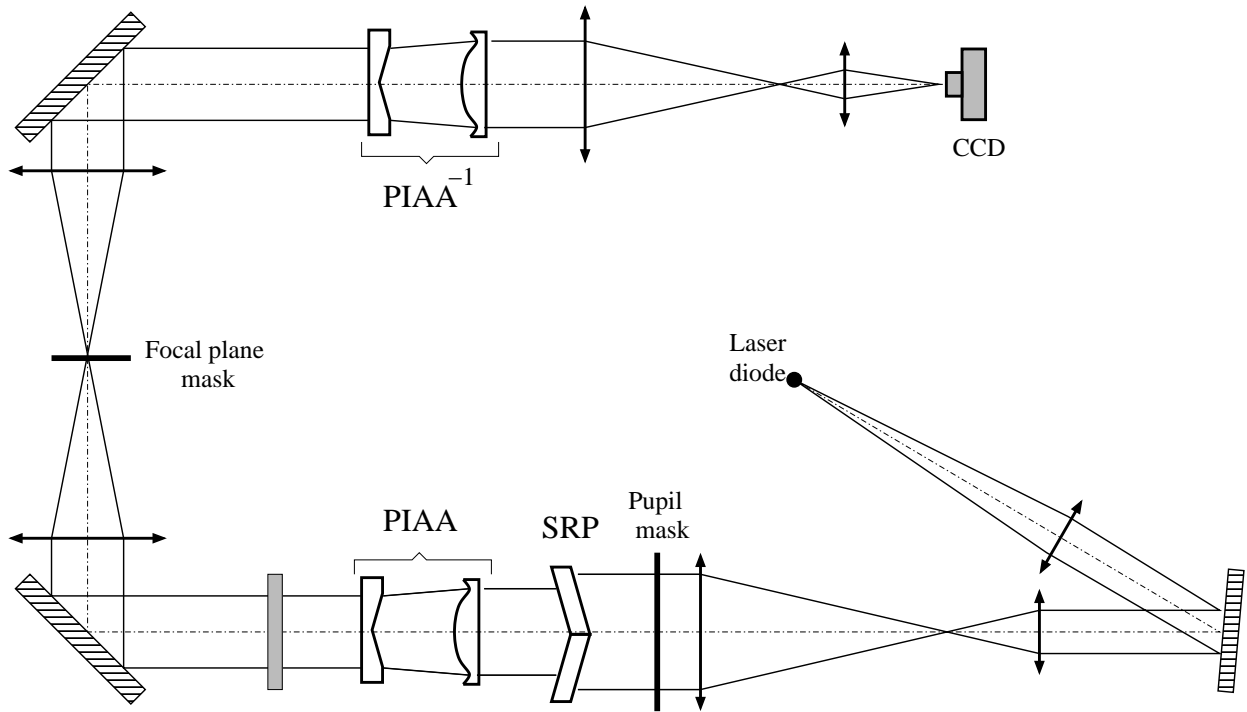}
\plotone{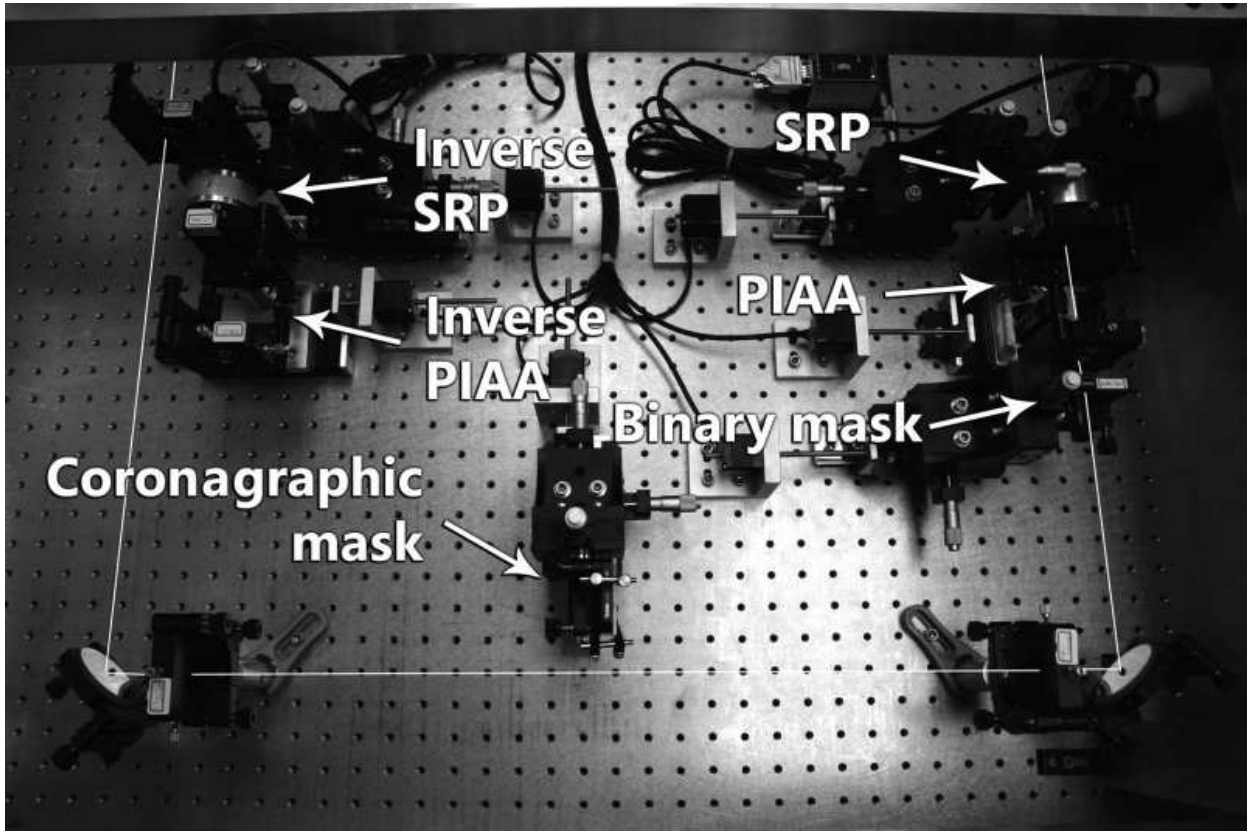}
  \caption{
    Top: optical layout used for the tests reported in this
    paper. Bottom: photo of the actual experiment. In the picture,
    the light is coming from the top right, goes into the SRP, the
    PIAA lenses and the binary mask. A lens focalizes the beam on the
    coronagraphic mask, and after there is the inverse PIAA and the
    inverse SRP. The beam then goes to the CCD camera.}
  \label{fig:experiment} 
\end{figure}

Except for the wavefront control system that is not tested in this
laboratory demonstration, the test bed is close to the system to be
installed onto the Nasmyth platform of the Subaru Telescope, after the
188-actuator AO system and before HiCIAO \citep{2008SPIE.7014E..42H}.
Figure \ref{fig:experiment} provides an overview of the system. On the
sketch, a 17 mm collimated beam comes horizontally from the
bottom-right corner, and goes through a mask simulating the Subaru
pupil. 
The beam then goes through the SRP (see \S\ref{sec:SRP}), the PIAA
(see \S\ref{sec:PIAA}) and the binary mask (cf. \S \ref{sec:bin_mask}).
At this point, the beam is suitably shaped for high-performance
coronagraphy.

The light reflects off a flat mirror and travels from bottom to top
through a first achromatic doublet that provides the first focal plane
where a coronagraphic mask is inserted, and another achromatic doublet
that provides another pseudo-pupil plane downstream.

The light then reflects off the top-left flat mirror and goes toward
the left, through an exact replica of the first PIAA system, only
mounted backwards. The motivation for this will be discussed in \S
\ref{sec:piaainv}. A two-doublet system makes an image on a CCD camera
that can be placed either in the pupil or the image plane.

Each element (c.f. bottom picture of Fig. \ref{fig:experiment}) is
mounted on a motorized translation stage to be removed from the
optical train in a few seconds. This provides good flexibility, and
permits the testing of multiple combinations for diagnostic, to the
extent that even if all ``active'' elements (SRP and PIAA) are taken
off the beam, the system still works as a perfectly functional
non-apodized Lyot coronagraph.

\subsection{Image acquisition and processing}

All acquisitions are made by a 1600x1200 pixels CCD camera (7.4 $\mu$m
square pixels). A optical density of 2 is used to avoid
saturation. Each image is cleaned by removing bias and hot
pixels, and for each simulation, we align and coadd 20 acquisitions to
reduce readout and photon noises.

The final two-doublet system mentioned in \S \ref{sec:layout} was
chosen to provide sampling better than Nyquist in the image plane,
while still allowing to image the pupil with a sufficient number of
pixels to observe the effect of the PIAA and the SRP. With these
``active'' elements out of the beam, the retained configuration
provides 2.3 pixels per $\lambda/D$ and 390 pixels in diameter for the
pupil of the telescope.

\subsection{Effect of the SRP}
\label{sec:srp_effect}

\begin{figure}
  \plotone{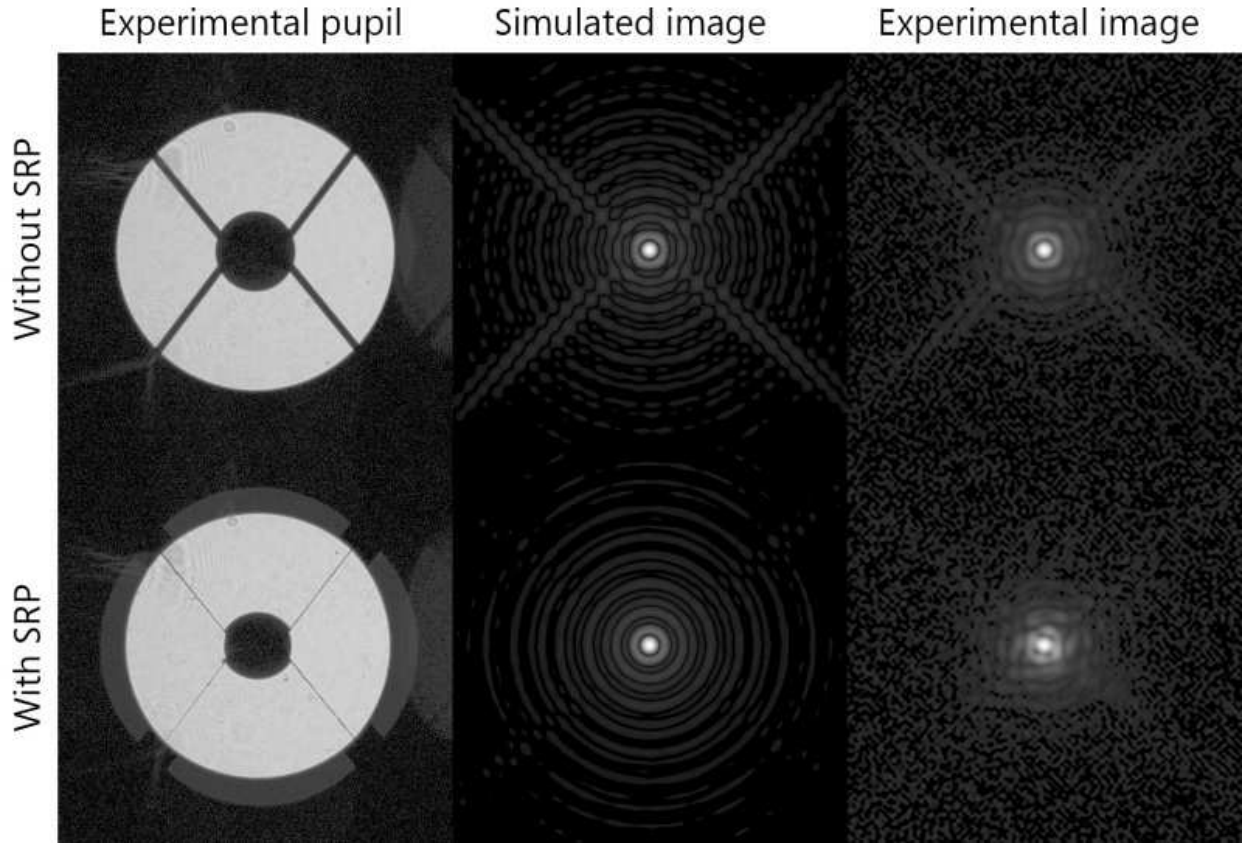}
  \caption{Effect of the SRP on the pupil. The three top panels show
    (from left to right): an image taken with the camera in a location
  conjugated with the pupil plane, showing the mask simulating the
  Subaru telescope pupil, as well as the corresponding simulated and
  experimental images. The three bottom panels show the same images
  when the SRP is inserted into the beam
  (cf. Sec. \ref{sec:srp_effect} for details).}
\label{fig:pupimwoandwsrp} 
\end{figure}

Figure \ref{fig:pupimwoandwsrp} demonstrates the effect of the SRP,
whose design and specifications are discussed in \S\ref{sec:SRP}. The
figure compares images of the pupil (left column) and the
corresponding PSFs (right column) taken with and without the SRP in
the beam.

The effect of the suppression of the spider vanes is clearly visible
as the diffraction spikes present in the standard imaging mode PSF
(top right image of Fig. \ref{fig:pupimwoandwsrp}) are diminished when
the SRP is introduced into the beam (bottom right image of same
figure).
Thinner spiders are however still present and contribute to weak
diffraction spikes in the images. This incomplete suppression of the
spiders can be attributed to the finite thickness of the glue holding
the quadrants together in the SRP.

Since no wavefront control system was used while conducting
these tests, nothing compensates the aberrations introduced by the
SRP, which beeing imperfect, necessarily deteriorates the quality of
the PSF. In order to offer a basis for comparison, the central column
of figure \ref{fig:pupimwoandwsrp} presents simulated versions of the
same images. Each PSF simulation was calculated by using the square
root of its corresponding pupil image for amplitude and constant
phases.
Without the SRP, a comparison of simulated and experimental images is
convincing evidence that the relaying lenses and mirrors are
sufficiently well-aligned not to introduce any obvious aberrations.

When inserting the SRP, the deterioration of the image quality is
non-negligible. Repeated measurements of a Strehl ratio $S = 0.48$
with the SRP ($S=0.98$ without) offer an estimate of the RMS aberration
it introduces through:

\begin{equation}
S = \exp{-(2\pi\sigma/\lambda)^2},
\label{eq:strehl}
\end{equation}

\noindent
which for $\lambda=633$ nm gives a RMS wavefront aberration $\sigma =
86$ nm. For reference, for $\lambda=1.6\mu$m, the same component would
provide a Strehl ratio $S = 0.89$.

This wavefront error, since it is static, does not have an impact on
the final PSF quality as long as (1) it is measured by the wavefront
sensor and (2) it is within the correction range of the deformable
mirror. The SCExAO architecture will include a wavefront sensor at the
science focal plane, and the SCExAO's DM stroke is about 1 $\mu$m. 
Since the remapping introduced by the SRP will turn a continuous
wavefront error into a sharp step, a DM located after the SRP cannot
easily correct such wavefront errors. The SCExAO's DM will therefore
be located before the SRP.

Because it is index-dependent (see equation \ref{eq:smalldelta}), the
displacement $\delta$ is a function of the wavelength and will go
from $\delta$ = 0.402 mm for $\lambda=1.6\mu$m to $\delta$ = 0.410 mm
for $\lambda=633$ nm.

We note that beeing optimized for the near-infrared, the coating of the
SRP is partially reflective (transmission T=0.89) at the test
wavelength. 
The pupil image with the SRP (Fig \ref{fig:pupimwoandwsrp}, lower
left) shows that each quadrant of the SRP creates a ghost toward the
outside of the pupil. These ghosts will be considerably fainter in the
near-IR wavelength for which the coatings have been designed." 

For now the tip-tilt of the SRP is controlled by two manual micrometer
drives, with an accuracy of a few arc seconds. In the final system,
higher accuracy remotely controlled actuators will be used.

\subsection{Pupil apodization}
\label{sec:apod}

\begin{figure}
  \plotone{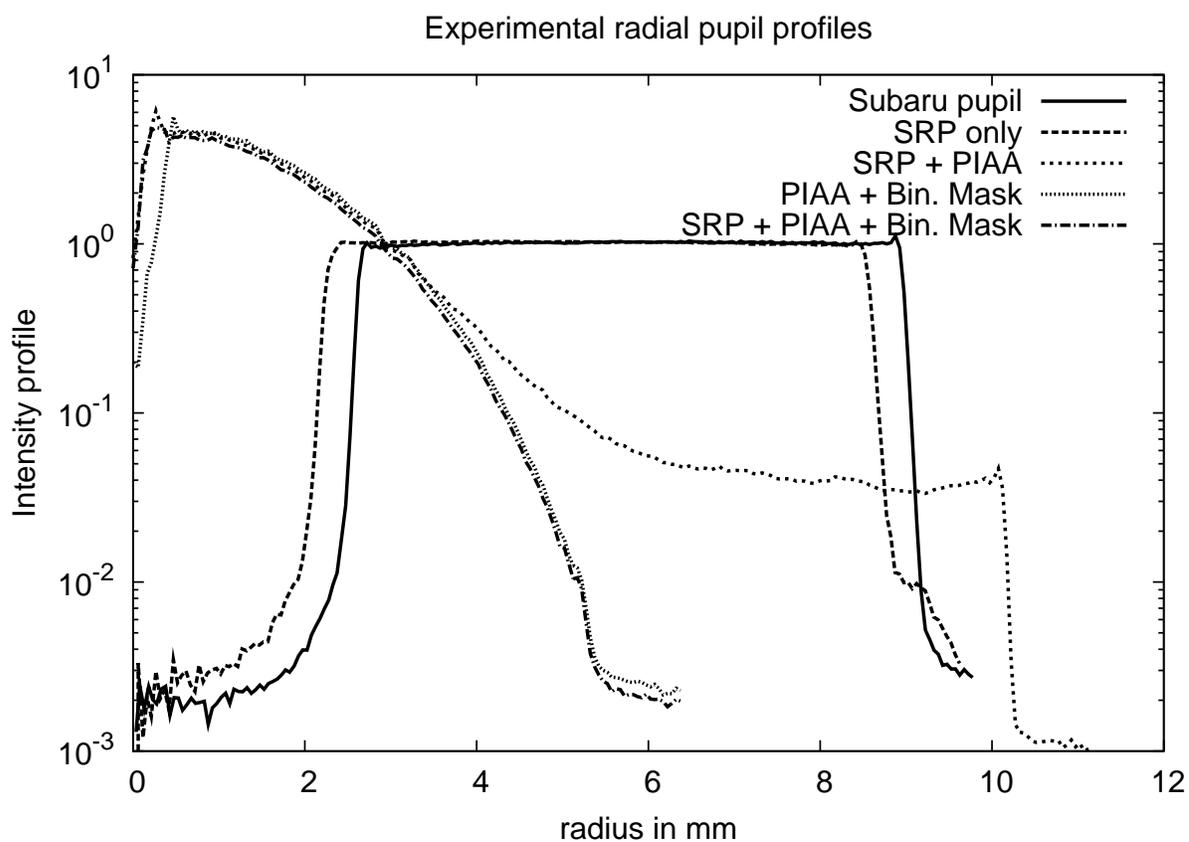}
  \caption{Radial intensity profiles of the pupil in the presence of
    different combinations of the active remapping elements: SRP, PIAA
    and binary mask presented in this paper. The mean illumination of
    the pupil in the absence of remapping (solid curve) is taken as
    reference. Note that unlike Fig. \ref{fig:apodesign}, the y-axis
    uses a logarithmic scale.}
  \label{fig:pupilprof} 
\end{figure}

The role of the apodization is to turn the hard-edge pupil,
responsible for the presence of rings in the PSF into the ideal
prolate spheroidal that will maximize the coronagraphic null.
In this system, the apodization is a three-step process that combines
the SRP, the PIAA and the binary mask.

Figure \ref{fig:pupilprof} plots pupil radial profiles measured with
our experiment using different combinations of these three elements,
to illustrate how each affects the pupil.

The effect of the SRP on the pupil profile is minimum. To fill up
the gap due to the spider vanes, it moves the four quadrants inwards
by 0.4 mm. The pupil beeing $\sim$ 9 mm in radius, this translates
into the 4.5 \% pupil size reduction observed when comparing the solid
and dashed line profiles in Fig. \ref{fig:pupilprof}. Note that the
central obscuration shrinks by the same amount.
Section \ref{sec:piaainv} will show how this apparent reduction of the
pupil diameter does not also translate into a penalty in angular
resolution.

The PIAA has a much more dramatic effect on the pupil. It takes the
light that was uniformly distributed in the original pupil and
concentrates its toward the center, filling up the void left by the
central obscuration.
The PIAA therefore turns the hard-edge (solid-line) profile into a
bright lobe (curve labeled SRP+PIAA on Fig. \ref{fig:pupilprof})
surrounded by a weak ($\sim$30 times fainter than originally),
slightly oversized uniform background.
The binary mask completes the apodization. It essentially substracts
the background (additional attenuation by a factor $\sim$10) and puts
the finishing touches to the central lobe, to maximize the rejection
by the occulting mask.

While looking at Fig. \ref{fig:pupilprof}, the reader may be under the
impression that the binary mask sabotages the hard work put in by the
PIAA and the SRP, by reducing the effective diameter of the pupil by a
factor of 2. The reader should be reassured by two things (1), the
vertical scale on the plots of Fig. \ref{fig:pupilprof} is logarithmic
and this apparent reduction of the pupil diameter is much less
significant than it appears and (2) the binary mask is located
downstream of the PIAA which leaves very little light on the edge of
the pupil, where the binary mask absorbs the most.
Overall, the transmission of the system is 92.3 \% excluding losses due
to coatings and beam clipping introduced to relax pupil alignment
tolerances. To achieve the same apodization profile on a
non-obstructed pupil, a CPA would have a 17.5 \% throughput.



\begin{figure}
  \plotone{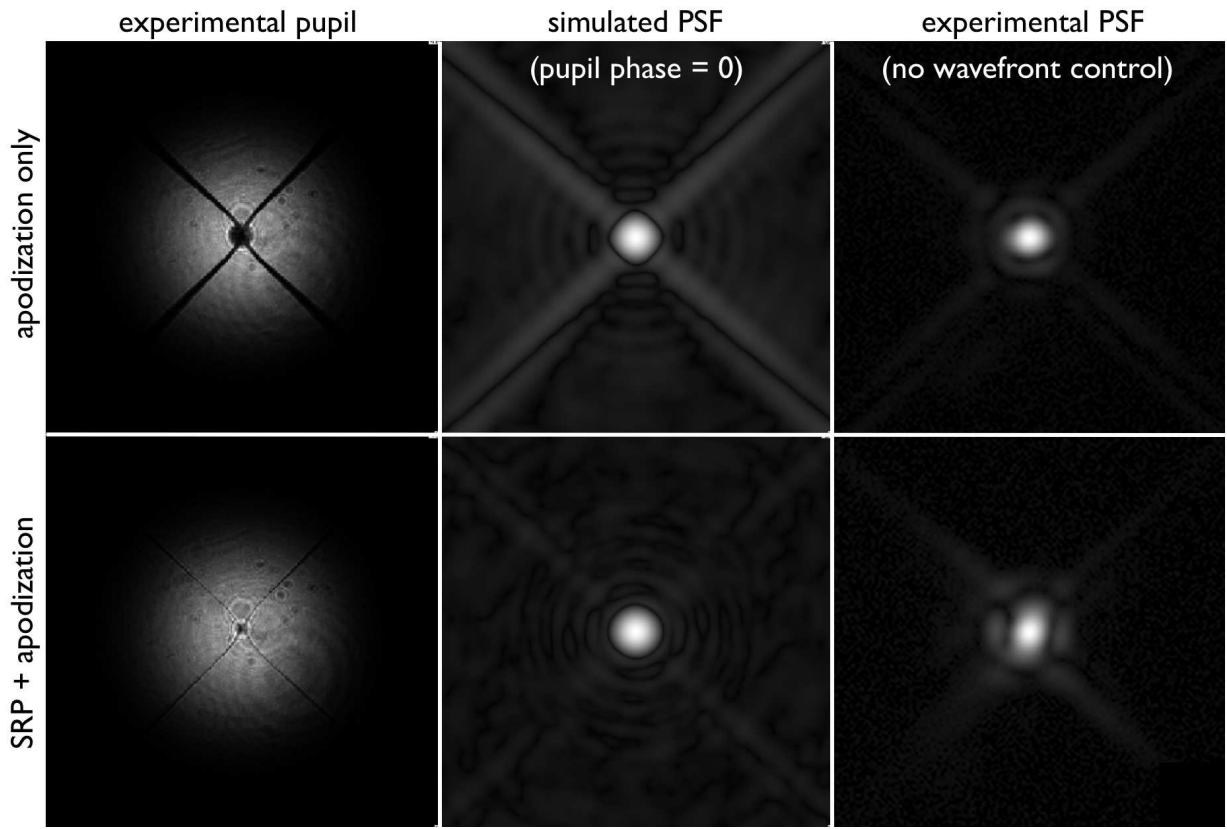}
  \caption{Apodization of the pupil: the theory and the experiment
    without and with SRP. Results in the image plane.}
  \label{fig:pupimwpiaa} 
\end{figure}

Figure~\ref{fig:pupimwpiaa} shows images of the illumination of the
apodized pupil taken with the SRP in and out of the beam, as well as
corresponding focal plane images.
The presence of the spider vanes in the top-left quadrant image offers
a way of visualizing the remapping of the pupil performed by the PIAA:
not only do the spider vanes get thinner toward the center of the
pupil as the distribution of complex amplitude gets concentrated,
because they do not exactly converge toward the geometric center, the
spider vanes are distorted by the remapping.
The images in the middle column show the corresponding simulated PSFs,
calculated from the experimental distribution of amplitude in the
pupil with a perfectly corrected wavefront (i.e. phase = 0). 
These images demonstrate the effect of the SRP as it reduces the
diffraction spikes of the spiders by a factor $10^3$.
The images in the right column are their experimentally obtained
counterparts. Although essential features of the PSFs are reproduced,
in the absence of wavefront control, the performance is obviously not
as good as the simulations predict. The elliptic (astigmatic)
experimental PSF obtained with the SRP can be attributed to an error
of alignment at the time of data acquisition.
Nevertheless, the fact that in both cases, these PSFs contain an
pseudo-Airy disk shows that the RMS of aberrations is below
$\lambda/4$, that is 150 nm. The DM the SCExAO Project will use
offers a peak-to-valley wavefront correction amplitude of 3
$\mu$m. The correction of this static 150-nm aberration will therefore
not represent a large fraction of the mirror stroke, which would
otherwise limit the SCExAO performance.

\section{Off-axis imaging performances}
\label{sec:offaxis}

It is a basic rule of conventional optics never to alter the pupil
if the basic object-image convolution relation is to be preserved. 
With components such as the SRP and the PIAA, the system presented in
this paper obviously violates this rule and this section explores the
consequences of their respective action on the field of view.

By design, SCExAO aims at complementing HiCIAO's parameter space
toward small angular separations. A strong constraint on the extent of
the extreme AO field of view on the actual system will come from the
DM used for the wavefront correction. Indeed, the finite number $N$ of
actuators projected across the pupil of the telescope sets the outer
working angle (OWA) of such a system to $(N/2)\times(\lambda/D)$.
The SCExAO DM offers at most $N=32$ such actuators. On the 8-meter
Subaru Telescope observing in H-band, this OWA is therefore limited to
0.6\arcsec ($16 \lambda/D$).

\subsection{SRP field of view: simulations and results}
\label{sec:SRP:fov}

\begin{figure}
  \plotone{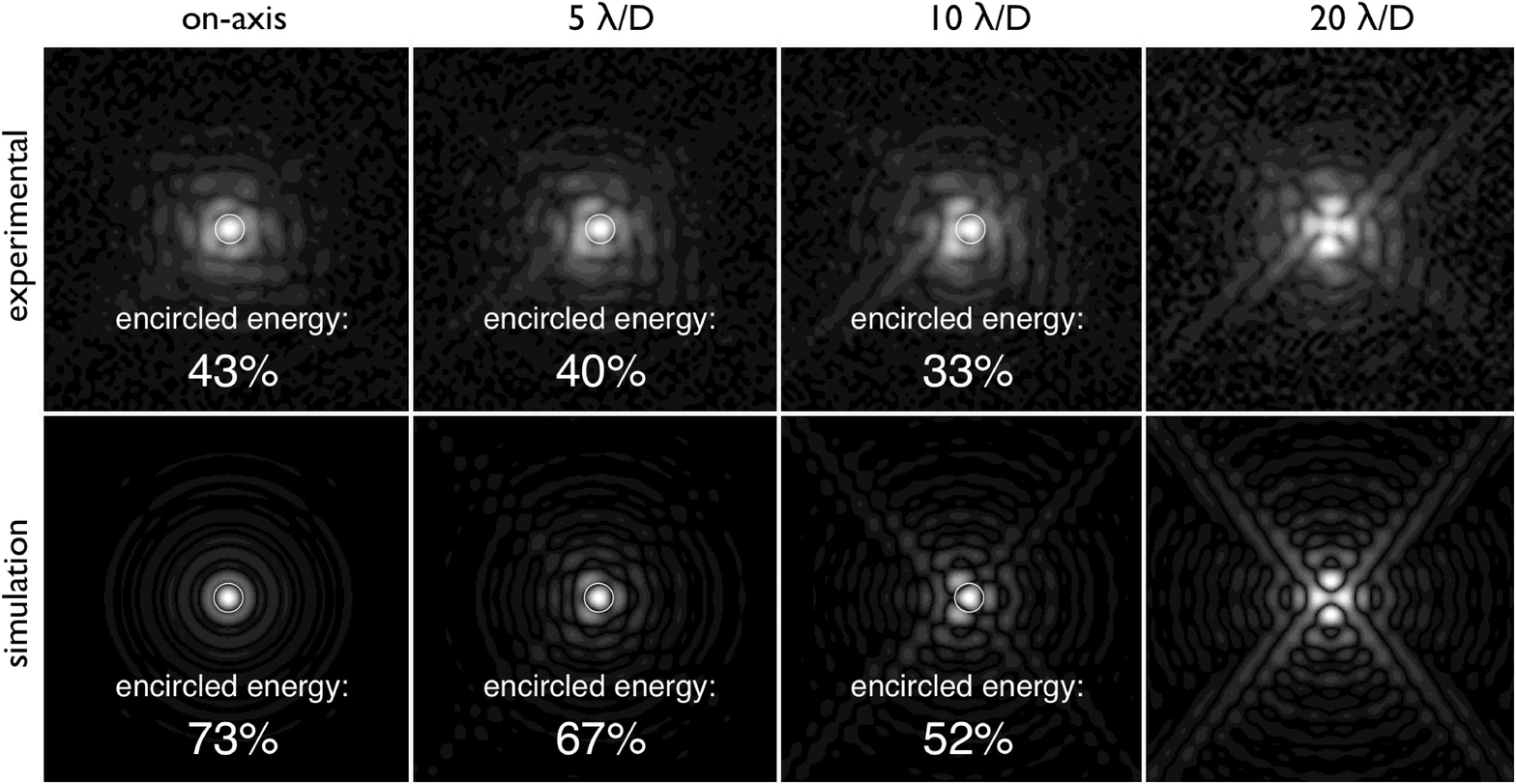}
  \caption{Comparison between simulated and experimental images of an
    off-axis source, from 0 to 20 $\lambda/D$ with a 5 $\lambda/D$
    step. The field of view of each image is $\pm 12 \lambda/D$.}
  \label{fig:psffieldsrp}
\end{figure}

Being fairly achromatic and inducing no light loss, the remapping of
the pupil performed by the SRP can be undone, after the coronagraphic
mask, by an inverse system, and two SRPs were manufactured in that
scope (cf. Fig. \ref{fig:experiment}).
After testing, one of the two however turned out of lesser quality,
which suggested to investigate how much the best SRP would alter the
``wide-field'' imaging capabilities of the SCExAO system.

Indeed, as explained in \S\ref{sec:SRP}, each quadrant of the SRP adds
an OPD that is a function of the tilt angle $\alpha$ for an on-axis
source (cf. Eqs. \ref{eq:opd} and \ref{eq:opd:dl}). The design however
guarantees, within tolerances, that the OPD is the same for all
quadrants, so that the wavefront of an on-axis source is continuous
after the SRP.

Although the tilt angle is the same, the azimuth of the tilt is
different for each quadrant. As a consequence, for an off-axis source,
the incidence angle will be different, resulting in a different OPD
for each quadrant. 
At a given wavelength $\lambda$, this will translate into a phase
offset $\Delta\phi$ that can be calculated using Eq. \ref{eq:opd:dl},
and substituting $\alpha$ by $\alpha+\theta$, with $\theta \ll \alpha$
(legitimate approximation since $\alpha = 5^\circ$):

\begin{equation}
  \Delta\phi = \frac{2\pi e (n-1)}{n \lambda} \alpha \cdot \theta.
\end{equation}

\noindent
The dependence on the tilt azimuth is expressed in the use of the dot
product $\alpha \cdot \theta$.
We have used this equation to predict how the off-axis affects the
quality of the PSF. Fig. \ref{fig:psffieldsrp} presents the result of
these predictions and compares them to laboratory images taken at
different off-axis angles with the SRP only.
The agreement between the simulations and the data over a fairly large
field of view (20 $\lambda/D$) is excellent and incidentally confirms
the validity of the proposed approximations.

One sees that for angular separations up to 10 $\lambda/D$, the SRP
produces an image with a pseudo-Airy disk. To better understand the
off-axis PSF degradation by the SRP, Fig.~\ref{fig:psffieldsrp} also
plots the energy encircled within the pseudo-Airy disk for the
relevant cases.
Theoretical values of this quantity for a non-obstructed and a 30 \%
obstructed telescope like Subaru are respectively 85 \% and 69 \%.
Because the SRP incidentally also reduces the size of the obscuration,
simulations show that on-axis, the encircled energy is 72 \%. It
degrades by an acceptable 7 \% for sources at a $5 \lambda/D$, and by
an unfortunate 30 \% at $10 \lambda/D$.
Experimental images and measurements conform to these simulations,
although the on-axis performance is reduced to 41 \%. This lesser
performance can be attributed to the absence of wavefront control in
this experiment.

Over a sufficiently large (10$\lambda/D$ in diameter) field of view,
the image quality appears very satisfactory, despite the absence of
wavefront control, whithin which high contrast imaging is performed:
an inverse SRP is not necessary in this observing mode.
For non-coronagraphic "wide field" observations, however, the SRP
should be removed from the beam.

\begin{figure}
  \plotone{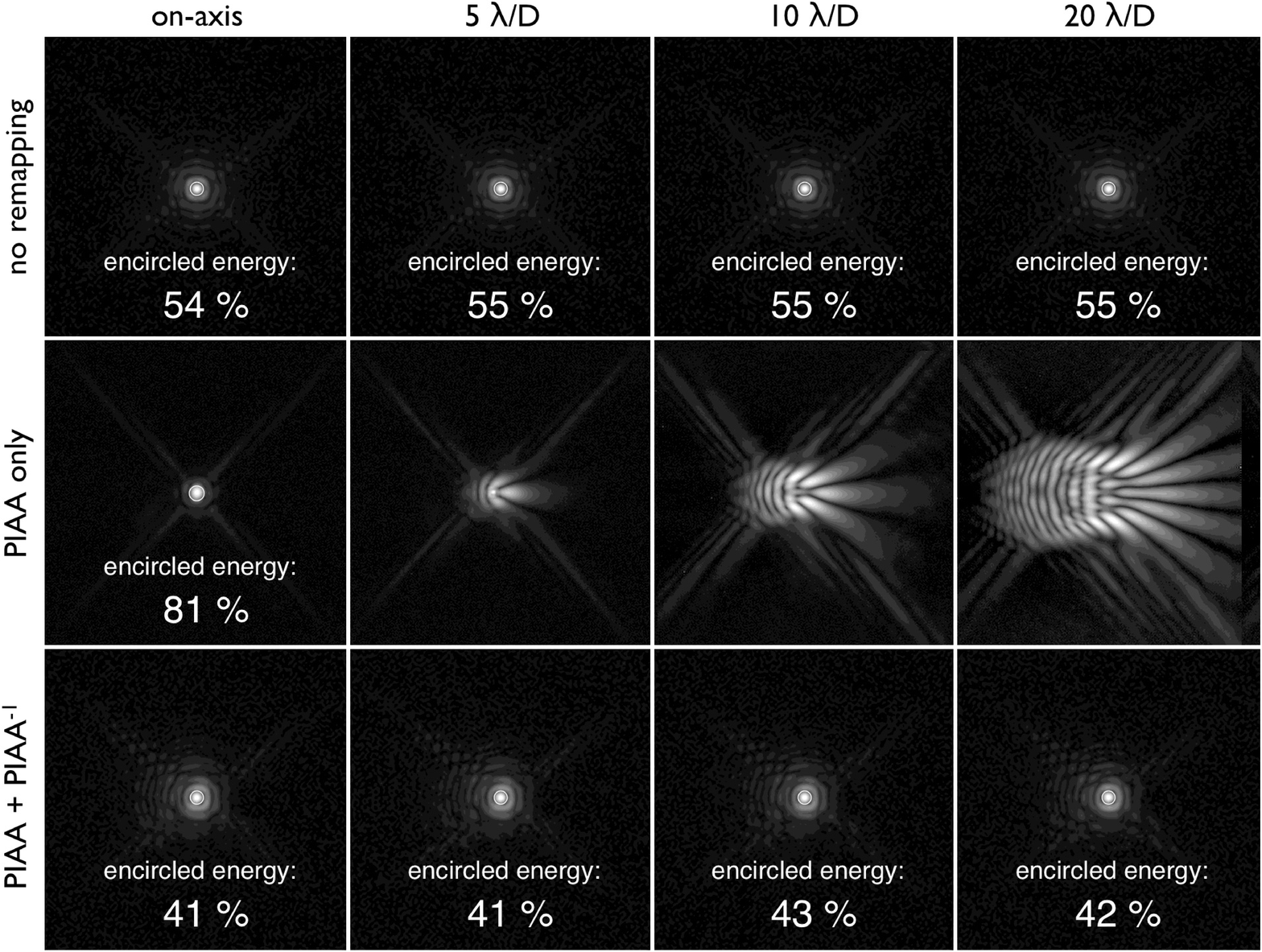}
  \caption{Experimental images of sources observed on and off-axis
    in the absence of SRP. The results of three configurations are
    shown. Top row: no apodization, middle row: PIAA + binary mask and
    bottom row: PIAA + binary mask + inverse PIAA.
    The huge off-axis aberrations the PIAA introduces are
    satisfactorily compensated by the inverse PIAA as the whole
    combination system provides diffraction limited images.}
  \label{fig:psffieldsrppiaa} 
\end{figure}

\subsection{Off-axis imaging with a PIAA}
\label{sec:piaainv}

For an off-axis source, the pupil-remapping the PIAA performs
introduces large aberrations to the wavefront, which translate into
very unsually shaped PSFs. The matter is not new and has been
extensively described by \citet{2005ApJ...622..744G} in the case of a
non-obstructed aperture.
The novelty in the PIAA that will equip the SCExAO system is that it
must accomodate the presence of a central obscuration, and as
presented in Section \ref{sec:apod}, the PIAA lenses used in
combination with the SRP manages to suppress it completely.
This however requires a somewhat brutal remapping, with dramatic
consequences on the PSF which finds itself pineapple-shaped at
separations greater than 10 $\lambda/D$
(cf. Fig. \ref{fig:psffieldsrppiaa}).

Fortunately, just like with any pupil remapping 
\citep{2002A&A...391..379G, 2003A&A...404..379G}, the aberrations the
PIAA introduces can entirely be corrected using an exact copy of the
PIAA, only plugged backwards, which restores the original pupil and
provides wide field of view imaging capabilities.
In practice, this so-called inverse PIAA is located in the pseudo
pupil plane located after the occulting mask (cf. Fig. 
\ref{fig:experiment}).
While this idea of using an inverse system to recover wide-field of
view capabilities after a PIAA coronagraph has often been mentioned in
PIAA related publications \citep{2003A&A...404..379G,
2005ApJ...622..744G, 2006ApJ...639.1129M, 2006ApJ...644.1246P}, this
paper reports the first successful implementation of such an inverse
system.

Figure \ref{fig:psffieldsrppiaa} shows a series of off-axis images
taken with and without the inverse PIAA and compares them to their
equivalent without any remapping at all (except for the SRP). These
images show that the inverse PIAA (bottom row) efficiently corrects
the huge aberrations the PIAA introduces in the first place (middle
row), and turns the pineapple-shaped off-axis images into more
conventional Airy-like images, virtually identical to the ones the
system produces with no beam apodization at all (top row).

Just as was done for the SRP (c.f. \S \ref{sec:SRP:fov}),
Fig.~\ref{fig:psffieldsrppiaa} also shows the energy encircled within
the pseudo-Airy disk of the PSFs (except for the pineapples).
The theoretical value of this quantity for the Subaru Telescope pupil
is 69 \%. In the absence of remapping (first row), the experimentally
measured encircled energy is 55 \%. Despite the absence of wavefront
control, the apodization by the PIAA manages to bring it a little over
80 \% on-axis.
With the inverse PIAA, the encircled energy stabilizes a little over
40 \%, whatever the angular position of the source. This 20 \% loss
(relative to the initial 55 \%) can be attributed to imperfection of
the optics. Nevertheless, this result, achieved in the absence of
wavefront control is remarkable: the obtention of an actual Airy disk
after the inverse PIAA shows that the wavefront residuals RMS is
better than $\lambda/4$. Observations in the near infrared and the
deformable mirror on the actual SCExAO system should significatively
improve this performance.

\section{Conclusion}

The Subaru Coronagraphic Extreme AO system is an instrument to be
installed onto the Subaru Telescope, as an upgrade to the recently
commissioned infrared camera HiCIAO \citep{2008SPIE.7014E..42H}.
The SCExAO system offers the means to probe the innermost part of
planetary systems by using technologies optimized for high contrast at
small angular separation.

Because it is constrained by the fundamental limits of diffraction and
the practical limits of actual telescopes, each high angular
resolution imaging system is the result of a multi-variable
optimization process: contrast versus angular separation versus field
of view.
While nothing fundamental prevents a PIAA-based coronagraph to achieve 
high contrast simultaneously with a small inner working angle and over
a very wide field of view, practical limits like a large central
obscuration and thick spider vanes, impose some level of
specialization for high performance.
The SRP+PIAA-based architecture of the SCExAO project was chosen to
open a unique niche of high contrast with a small inner working
angle for an 8-meter telescope.
While focusing on a 10 $\lambda/D$-diameter field of view may seem
restrictive, the recent results of other high angular resolution
techniques working on a very small field of view, like aperture
masking interferometry \citep{2009ApJ...695.1183M} which SCExAO will
also implement, show that such specialization pays off.

In that scope, this paper introduces two critical components of the
system: a new design for the PIAA apodizer that accomodates the
presence of a central obscuration and a compound phase plate that
suppresses the spider vanes present in most modern telescope designs.
The intensive tests conducted on these advanced optical components, at
a wavelength three times shorter than nominal, demonstrate
satisfactory performance despite the absence of wavefront control.
We are now integrating these optics with the wavefront control system
necessary to produce high contrast images.

The SCExAO system, in its first installment, is designed to provide
high contrast imaging in the 1 to 4 $\lambda/D$ separation range in
imaging mode only (no spectroscopy). 
By using achromatic near-IR lenses for relay optics and a low
dispersion material (CaF2) for the PIAA lenses, chromatic effects in
SCExAO are mitigated. Polychromatic operation of SCExAO (for
spectroscopy) has however not been explored at this point, and will
likely require redesign of several components in the optical train
(most importantly to compensate for the linear dependancy of the
diffraction limit $\lambda/D$ with wavelength in the focal plane).
The effect of SCExAO residual chromatic effects on Simultaneous
Differential Imaging (SDI) has also not been evaluated, but is
believed to be manageable over the $\sim 1 \%$ wide spectral band
covered by each of the SDI filter. We note that SCExAO does not rely
on SDI for differential detection: thanks to the high order SCExAO DM,
coherence-based techniques to separate residual speckles from true
sources (as demonstrated in \cite{guyon09}) will be the the primary
speckle calibration mode.


\begin{thebibliography}{0}
\expandafter\ifx\csname natexlab\endcsname\relax\def\natexlab#1{#1}\fi

\end{thebibliography}


\begin{thebibliography}{30}
\expandafter\ifx\csname natexlab\endcsname\relax\def\natexlab#1{#1}\fi

\bibitem[{Abe} et~al.(2006){Abe}, {Murakami}, {Nishikawa} \&
  {Tamura}]{2006A&A...451..363A}
{Abe} L., {Murakami} N., {Nishikawa} J., {Tamura} M., 2006, \aap, 451, 363

\bibitem[{Aime} \& {Soummer}(2003)]{2003EAS.....8...79A}
{Aime} C., {Soummer} R., 2003, in { EAS Publications Series\/}, edited by
  C.~{Aime}, R.~{Soummer}, vol.~8 of { EAS Publications Series\/},  79--92

\bibitem[{Beuzit} et~al.(2006){Beuzit}, {Feldt}, {Dohlen}
  et~al.]{2006Msngr.125...29B}
{Beuzit} J.-L., {Feldt} M., {Dohlen} K., et~al., 2006, The Messenger, 125, 29

\bibitem[{Biller} et~al.(2007){Biller}, {Close}, {Masciadri}
  et~al.]{2007ApJS..173..143B}
{Biller} B.~A., {Close} L.~M., {Masciadri} E., et~al., 2007, \apjs, 173, 143

\bibitem[{Cavarroc} et~al.(2006){Cavarroc}, {Boccaletti}, {Baudoz}, {Fusco} \&
  {Rouan}]{2006A&A...447..397C}
{Cavarroc} C., {Boccaletti} A., {Baudoz} P., {Fusco} T., {Rouan} D., 2006,
  \aap, 447, 397

\bibitem[{Guyon}(2003)]{2003A&A...404..379G}
{Guyon} O., 2003, \aap, 404, 379

\bibitem[{Guyon}(2005)]{2005ApJ...629..592G}
{Guyon} O., 2005, \apj, 629, 592

\bibitem[{Guyon} et~al.(2005){Guyon}, {Pluzhnik}, {Galicher}, {Martinache},
  {Ridgway} \& {Woodruff}]{2005ApJ...622..744G}
{Guyon} O., {Pluzhnik} E.~A., {Galicher} R., {Martinache} F., {Ridgway} S.~T.,
  {Woodruff} R.~A., 2005, \apj, 622, 744

\bibitem[{Guyon} et~al.(2006){Guyon}, {Pluzhnik}, {Kuchner}, {Collins} \&
  {Ridgway}]{2006ApJS..167...81G}
{Guyon} O., {Pluzhnik} E.~A., {Kuchner} M.~J., {Collins} B., {Ridgway} S.~T.,
  2006, \apjs, 167, 81

\bibitem[{Guyon} et~al.(2009){Guyon}, {Pluzhnik}, {Martinache} et~al.]{guyon09}
{Guyon} O., {Pluzhnik} E.~A., {Martinache} F., et~al., 2009, \pasp, submitted

\bibitem[{Guyon} \& {Roddier}(2002)]{2002A&A...391..379G}
{Guyon} O., {Roddier} F., 2002, \aap, 391, 379

\bibitem[{Hodapp} et~al.(2008){Hodapp}, {Suzuki}, {Tamura}
  et~al.]{2008SPIE.7014E..42H}
{Hodapp} K.~W., {Suzuki} R., {Tamura} M., et~al., 2008, in { Society of
  Photo-Optical Instrumentation Engineers (SPIE) Conference Series\/}, vol.
  7014 of { Society of Photo-Optical Instrumentation Engineers (SPIE)
  Conference Series\/}

\bibitem[{Kalas} et~al.(2008){Kalas}, {Graham}, {Chiang}
  et~al.]{2008Sci...322.1345K}
{Kalas} P., {Graham} J.~R., {Chiang} E., et~al., 2008, Science, 322, 1345

\bibitem[{Kasdin} et~al.(2005){Kasdin}, {Vanderbei}, {Littman} \&
  {Spergel}]{2005ApOpt..44.1117K}
{Kasdin} N.~J., {Vanderbei} R.~J., {Littman} M.~G., {Spergel} D.~N., 2005, \ao,
  44, 1117

\bibitem[{Kasdin} et~al.(2003){Kasdin}, {Vanderbei}, {Spergel} \&
  {Littman}]{2003ApJ...582.1147K}
{Kasdin} N.~J., {Vanderbei} R.~J., {Spergel} D.~N., {Littman} M.~G., 2003,
  \apj, 582, 1147

\bibitem[{Kasper} et~al.(2007){Kasper}, {Apai}, {Janson} \&
  {Brandner}]{2007A&A...472..321K}
{Kasper} M., {Apai} D., {Janson} M., {Brandner} W., 2007, \aap, 472, 321

\bibitem[{Labeyrie}(1996)]{1996A&AS..118..517L}
{Labeyrie} A., 1996, \aaps, 118, 517

\bibitem[{Lafreni{\`e}re} et~al.(2007){Lafreni{\`e}re}, {Doyon}, {Marois}
  et~al.]{2007ApJ...670.1367L}
{Lafreni{\`e}re} D., {Doyon} R., {Marois} C., et~al., 2007, \apj, 670, 1367

\bibitem[{Lagrange} et~al.(2008){Lagrange}, {Gratadour}, {Chauvin}
  et~al.]{2008arXiv0811.3583L}
{Lagrange} A.~., {Gratadour} D., {Chauvin} G., et~al., 2008, ArXiv e-prints

\bibitem[{Lyot}(1939)]{1939MNRAS..99..580L}
{Lyot} B., 1939, \mnras, 99, 580

\bibitem[{Macintosh} et~al.(2008){Macintosh}, {Graham}, {Palmer}
  et~al.]{2008SPIE.7015E..31M}
{Macintosh} B.~A., {Graham} J.~R., {Palmer} D.~W., et~al., 2008, in { Society
  of Photo-Optical Instrumentation Engineers (SPIE) Conference Series\/}, vol.
  7015 of { Society of Photo-Optical Instrumentation Engineers (SPIE)
  Conference Series\/}

\bibitem[{Marois} et~al.(2008){Marois}, {Macintosh}, {Barman}
  et~al.]{2008arXiv0811.2606M}
{Marois} C., {Macintosh} B., {Barman} T., et~al., 2008, ArXiv e-prints

\bibitem[{Martinache} et~al.(2006){Martinache}, {Guyon}, {Pluzhnik}, {Galicher}
  \& {Ridgway}]{2006ApJ...639.1129M}
{Martinache} F., {Guyon} O., {Pluzhnik} E.~A., {Galicher} R., {Ridgway} S.~T.,
  2006, \apj, 639, 1129

\bibitem[{Martinache} et al.(2009){Martinache}, {Rojas-Ayala},
  {Ireland}, {Lloyd} \& {Tuthill}]{2009ApJ...695.1183M}
{Martinache} F., {Rojas-Ayala} B., {Ireland} M.~J., {Lloyd} J.~P., 
{Tuthill} P.~G., 2009, \apj, 695, 1183 


\bibitem[{Martinez} et~al.(2008){Martinez}, {Boccaletti}, {Kasper}
  et~al.]{2008A&A...492..289M}
{Martinez} P., {Boccaletti} A., {Kasper} M., et~al., 2008, \aap, 492, 289

\bibitem[{Pluzhnik} et~al.(2006){Pluzhnik}, {Guyon}, {Ridgway}
  et~al.]{2006ApJ...644.1246P}
{Pluzhnik} E.~A., {Guyon} O., {Ridgway} S.~T., et~al., 2006, \apj, 644, 1246

\bibitem[{Racine} et~al.(1999){Racine}, {Walker}, {Nadeau}, {Doyon} \&
  {Marois}]{1999PASP..111..587R}
{Racine} R., {Walker} G.~A.~H., {Nadeau} D., {Doyon} R., {Marois} C., 1999,
  \pasp, 111, 587

\bibitem[{Roddier} \& {Roddier}(1997)]{1997PASP..109..815R}
{Roddier} F., {Roddier} C., 1997, \pasp, 109, 815

\bibitem[{Sivaramakrishnan} \& {Lloyd}(2005)]{2005ApJ...633..528S}
{Sivaramakrishnan} A., {Lloyd} J.~P., 2005, \apj, 633, 528

\bibitem[{Soummer}(2005)]{2005ApJ...618L.161S}
{Soummer} R., 2005, \apjl, 618, L161

\bibitem[{Soummer} et~al.(2003){Soummer}, {Aime} \&
  {Falloon}]{2003A&A...397.1161S}
{Soummer} R., {Aime} C., {Falloon} P.~E., 2003, \aap, 397, 1161

\end{thebibliography}

\end{document}